\documentclass[prd,aps,amssymb,nofootinbib,superscriptaddress,floatfix]{revtex4}

\usepackage{amsmath}
\usepackage{graphicx}
\usepackage{latexsym}
\usepackage{amsfonts}
\usepackage{url,hyperref}
\usepackage{bm}
\usepackage{color}

\newcommand{\nn}{\nonumber\\}
\newcommand{\beq}{\begin{equation}}
\newcommand{\eeq}{\end{equation}}
\newcommand{\bed}{\begin{displaymath}}
\newcommand{\eed}{\end{displaymath}}
\def\bea{\begin{eqnarray}}
\def\eea{\end{eqnarray}}

\begin{document}

\title{Graviton emission from simply rotating Kerr-de Sitter black holes: \\
Transverse traceless tensor graviton modes}
\author{Jason~Doukas}
\email[Email: ]{j.doukas@ms.unimelb.edu.au}
\affiliation{Department of Mathematics and Statistics, The University of Melbourne, Parkville, Victoria 3010, Australia.}
\author{H.~T.~Cho}
\email[Email: ]{htcho@mail.tku.edu.tw}
\affiliation{Department of Physics, Tamkang University, Tamsui, Taipei, Taiwan, Republic of China}
\author{A.~S.~Cornell}
\email[Email: ]{alan.cornell@wits.ac.za}
\affiliation{National Institute for Theoretical Physics; School of Physics, University of the Witwatersrand, Wits 2050, South Africa}
\author{Wade~Naylor}
\email[Email: ]{naylor@se.ritsumei.ac.jp}
\affiliation{Department of Physics, Ritsumeikan University, Kusatsu, Shiga 525-8577, Japan}

\begin{abstract}
In this article we present results for tensor graviton modes (in seven dimensions and greater, $n\geq 3$) for greybody factors of Kerr-dS black holes and for Hawking radiation from simply rotating $(n+4)$-dimensional Kerr black holes. Although there is some subtlety with defining the Hawking temperature of a Kerr-dS black hole, we present some preliminary results for emissions assuming the standard Hawking normalization and a Bousso-Hawking-like normalization. 
\end{abstract}

\date{\today}
\maketitle


\section{Introduction}

\par The reduction of the graviton perturbation equations into master variable equations for higher-dimensional  black holes has been one of the great challenges in recent years, for example see the review in reference \cite{Kodama:2007ph}. In the case of spherically symmetric static spacetimes, master equations for the scalar, vector and tensor decompositions of the metric perturbations have, however, been achieved \cite{Kodama:2003jz}. The method based upon the gauge invariant formalism, developed in reference \cite{Kodama:2000fa} has also allowed for the separation of the tensor mode decomposition of simply rotating Myers-Perry-(A)dS black holes \cite{Kodama:2007ph,Kodama:2007sf}, which has recently been used for a stability analysis of Kerr-AdS black holes \cite{Kodama:2009rq,Kodama:2009bf}.

\par The interest in Hawking radiation has stemmed from the idea that mini-TeV black holes might be created at the Large Hadron Collider (LHC) or in cosmic ray showers in the upper atmosphere of the Earth \cite{Landsberg:2001sj,Emparan:2001kf,Kanti:2004nr}. A likely signal would be the Hawking emission of particles from such TeV black holes as simulated by the BlackMax event generator \cite{Dai:2007ki}, where only the known greybody factors at that time were included. With that aim in mind a numerous amount of research has investigated Hawking radiation from brane, bulk and tense-brane black holes, for a recent review see reference \cite{Kanti:2009sz} and references therein. One of the most important issues is that of bulk graviton emissions \cite{Cornell:2005ux,Creek:2006ia,Cardoso:2005mh,Cardoso:2005vb}, as the hypotheses that black holes mainly radiate on the brane \cite{Emparan:2000rs} needs to be verified \cite{Cavaglia:2003hg}. For Schwarzschild \cite{Cardoso:2005mh}, Schwarzschild-dS \cite{Wu:2008rb} and Tense-Schwarzschild \cite{Dai:2006hf} black holes the consensus is that emissions are brane dominated. 

\par The same question should be addressed for rotating black holes, and some progress in this direction has been made for special cases where the maximal number of rotation parameters are equal (that is, degenerate) \cite{Kunduri:2006qa,Murata:2008yx}. However, for the case of simply rotating (single rotation plane) Myers-Perry-(A)dS black holes, nothing has yet been said. As we mentioned, only tensor modes are currently possible, which is the purpose of this article. Even for spin-$0$ fields there appears to be no work on greybody factors in Kerr-(A)dS spacetimes, for example see reference \cite{Creek:2007pw} for simply rotating Kerr black holes in $(n+4)$-dimensions and reference \cite{Kobayashi:2007zi} for the tense-brane variant (for the Schwarzschild-(A)dS case, see reference \cite{Harmark:2007jy}).

\par In general there are $N=[(n+3)/2]$ rotation parameters for higher dimensional Kerr-(A)dS black holes \cite{Gibbons:2004uw} and work for multiple rotation parameters is in progress (also see reference \cite{Nomura:2005mw}); however, a simple separation into a four dimensional base space and line element for the unit $n$-sphere appears difficult to obtain (although see reference \cite{Vasudevan:2004ca}). Thus, in this work we shall focus on a simply rotating black hole in $(n+4)$-dimensional Kerr-de Sitter spacetime, which was first derived rigorously in reference \cite{Gibbons:2004uw}, also see reference \cite{Hawking:1998kw}. Due to problems with the definition of a vacuum state for a black hole in a box \cite{Winstanley:2001nx}, we shall not consider the Kerr-AdS case here. We should also mention that much of the analysis here will only apply once   stability of Kerr-dS black holes (using QNM analysis) has been performed.

\par The set up of the paper is as follows. In the next section we give the background field equations and discuss the separation of the tensor graviton wave equation. In section III we discuss how to evaluate the eigenvalue for the angular separation constant on a Kerr-dS spacetime. In section IV results for the greybody factors calculated via a numerical matching method are presented, with results on superradiance given in section V. In section VI some issues relating to the definition of surface gravity in asymptotically de Sitter spacetimes are discussed. In section VII results for the energy and angular emission rates are presented, while in the last section we close the article with conclusions.


\section{Kerr-(A)dS Background}

\par The metric for the $(n+4)$-dimensional (A)dS-Kerr black hole for the simply rotating case \cite{Gibbons:2004uw}, when written in Boyer-Lindquist coordinates, is \cite{Kodama:2009rq}:
\beq
ds^2 = -\frac{\Delta_r}{\rho^2}\Biglb(dt-\frac{a}{1+\lambda a^2}\sin^2\theta d\phi\Bigrb)^2
+\frac{\Delta_\theta\sin^2\theta}{\rho^2}\Biglb(adt-\frac{r^2+a^2}{1+\lambda a^2}d\phi\Bigrb)^2
  + \frac{\rho^2}{\Delta_r}dr^2+ \frac{\rho^2}{\Delta_\theta}d\theta^2
   +r^2\cos^2\theta d\Omega_{n}^2~,
   \label{metric}
\eeq
where for positive cosmological constant $\Lambda$ we have
\beq
\lambda = {2 \Lambda \over (n+2)(n+3)} 
\eeq
(in the above we have set Newton's $(n+4)$-dimensional constant to unity). $d\Omega_n^2$ is the metric of the $n$-dimensional unit sphere, and
\beq
\Delta_r= (r^2+a^2)\biglb(1- \lambda r^2\bigrb) - \frac{2M}{r^{n-1}}, \qquad   \Delta_\theta=1+\lambda  a^2 \cos^2\theta ,\qquad
\rho^2= r^2 + a^2\cos^2\theta~.
\eeq

\par The wave equation for the tensorial mode of the gravitational perturbation for $n\geq 3$ is equivalent to the wave equation of a massless free scalar field \cite{Kodama:2007ph,Kodama:2009rq}, defined by 
\beq
{1\over \sqrt{-g} } \partial_\mu (\sqrt{-g} g^{\mu\nu} \partial_\nu \Phi)=0~,
\eeq
where the determinant is given by the product of the base metric \cite{Kodama:2009rq} and higher-dimensional spherical harmonics \cite{Casals:2008pq}
\beq
\sqrt{-g} =|1+\lambda a^2|^{-1} \rho^2 \sin \theta~\otimes~ r^n\cos^n\theta~ \prod_{i=1}^{n-1} \sin^i \theta_i~.
\eeq
The separation of the wave equation is implemented by making the ansatz:
\begin{eqnarray}
\Phi = e^{i\omega t-im\varphi} R(r) S_{jlm}(\theta)Y_{j,i_1,i_2,\ldots,i_{n-1}}(\theta_{n-1},\phi)~,
\end{eqnarray}
where $Y_{j,i_1,i_2,\ldots,i_{n-1}}(\theta_{n-1},\phi) $ are the hyperspherical harmonics on the $n$-sphere with eigenvalues $-j(j+n-1)$. This separation ansatz leads to a generalized hyper-spheroidal equation for the $S_{jlm}(\theta)$ functions, given by:
\beq
{1\over\sin\theta \cos^n\theta  } \partial _\theta \left((1+\lambda a^2\cos^2\theta) \sin\theta \cos^n\theta \partial_\theta S\right)+\left(A_{ljm} -\frac{m^2(1+\lambda a^2)}{\sin^2\theta} -\frac{a^2\omega^2\sin^2\theta}{1+\lambda a^2\cos^2\theta} -\frac{j(j+n-1)}{\cos^2\theta} \right) S=0
\label{angular}
\eeq
and an equation for $R(r)$, the radial equation, which satisfies
\beq
{1\over r^n} \partial_r\left(r^n\Delta_r \partial_r R\right)
+\bigglb[ -A_{ljm}
   +\frac{a^2 m^2}{\Delta_r}
    \bigglb( (1+\lambda a^2)(1-{\lambda} r^2)
               -\frac{2\lambda M}{ r^{n-1}}  \biggrb)
    -\frac{4Ma m\omega}{r^{n-1}\Delta_r}
    +\frac{(r^2+a^2)^2\omega^2}{\Delta_r}
    -\frac{j(j+n-1)a^2}{r^2} \biggrb] R=0~,
\label{radial}
\eeq
where $A_{ljm}$ is the separation constant. These equations are coupled via $\omega$, in contrast to the static case where the $\omega^2$ term appears in the angular equation only. The restrictions on $m$, $j$ and $l$ \cite{Berti:2005gp} are:
\beq
l>j +|m|~; \quad  \frac{l- j +|m|}{2}\in \{0,1,2...\}~,
\eeq
with $j=2,3,\dots,l$ and $|m|=0,1,\dots, l-j$. The degeneracy for a traceless symmetric tensor on an $n$-sphere \cite{Rubin:1984tc} is given by:   
\beq
D_{j}^T=\frac{(n+1)(n-2)(j+n)(j-1)(2j+n-1)(j+n-3)!}{2(n-1)!(j+1)!}~.
\label{deg}
\eeq 
Equations (\ref{angular}) and (\ref{radial}) reduce to those for a free scalar field on an asymptotically flat background, $\lambda \to 0$, after an eigenvalue shift \cite{Creek:2007pw}. In section \ref{sec:eval} we numerically solve for the angular eigenvalue in equation (\ref{angular}) using some of the methods discussed in our recent work for angular eigenvalues of Kerr-(A)dS black holes \cite{Cho:2009wf}. 


\subsection{Horizons}

\par For the simply rotating Kerr-dS black hole the metric is stationary and independent of $t$ and $\phi$, and hence there are two Killing vectors $K_t^a =(1,0,0,0)$ and $K_\phi^a=(0,1,0,0)$ (coming from the base metric coordinates: $(t,\phi,r,\theta)$).  Following the approach used in reference \cite{Gibbons:2004uw} the horizons coincide with an orbit of the Killing vector field:
\beq
K= \gamma_t {\partial \over \partial_t } + \gamma_\phi \Omega(r_h) {\partial \over \partial_\phi }~, \qquad\qquad \Omega(r) = {a (1-\lambda r^2)\over a^2+ r^2}~,
\eeq
 where the surface gravity, $\kappa$, is related by 
\beq
K^a \nabla_a K_b = \kappa K_b~.
\eeq
Note that we have included the normalisations $\gamma_t,\,\gamma_\phi$ for each respective Killing vector (in Kerr-AF spacetimes these normalizations can be chosen by choosing a Killing field that goes to unit time and axial ($m^2/r^2\to 1$) at infinity). For a vector satisfying the Killing condition we can rewrite the above equation as \cite{Gibbons:2004js,Gibbons:2004uw}:
\beq
\frac 1 2 \nabla_b L^2 = \kappa K_b~,
\label{killsurf}
\eeq
where
\beq
L^2(r,\theta) \equiv-K^a K_a = -\gamma^2 ( g_{tt} + 2\Omega_h g_{\phi t} + \Omega_h^2 g_{\phi\phi} )~,
\label{Lsq}
\eeq
and we have set $\gamma_t=\gamma_\phi=\gamma$ (which ensures that the angular velocity of the horizon, $\Omega_h$, is independent of the normalisation). For example, 
\beq
K^2_t = \gamma^2 g_{tt} =- \frac{\gamma^2 \Delta_r}{(1+\lambda a^2)^2\rho^2}\left[(1+\lambda a^2)(1-\lambda r^2)\rho^2
-\frac{2M}{r^{n-1}}\Delta_\theta\right]~, 
\eeq
with similar expressions for $g_{t\phi}$ and $g_{\phi\phi}$. Furthermore, from the properties of the two Killing vectors it is also possible to derive the constraint (see reference \cite{Townsend:1997ku})
\beq
(K_t \cdot K_\phi)^2-K_t^2 K_\phi^2 =\gamma^2 g_{t\phi}^2- \gamma^2 g_{tt} g_{\phi\phi}=   {\gamma^2  \Delta_r \Delta_\theta \over 1+\lambda a^2} \sin^2\theta=0~,
\eeq
which implies the location of the horizons: $\Delta_r=0$.  For dimensions $n\geq 3$ this always has two positive roots if $\lambda > 0$ and hence we can  parameterize the black hole mass, $M$,  in terms of the horizon radius $r_h$:
\beq 
\label{eqn:M}
2M = r_h^{n-1} (r_h^2+a^2) (1- \lambda r_{h}^2 )~.
\eeq


\subsection{WKBJ Form}

\par The radial equation \eqref{radial} can be put into WKBJ form by defining the transform:
\beq
R(r)= r^{-n/2} (r^2+a^2)^{-1/2} \Phi(r)~,
\eeq
and tortoise coordinates \cite{Kodama:2009rq}:
\beq
dy={r^2 +a^2 \over \Delta_r }dr~,
\label{tort}
\eeq
where
\beq
{\Delta_r \over r^2 +a^2}= 1 - {2M \over (r^2+a^2 ) r^{n-1}}-{2 \Lambda \over (n+2)(n+3)}  r^2 ~.
\eeq
After defining the dimensionless variables: $x= r/r_h$, $\omega_{\star} = \omega r_h$, $y_{\star}= y/r_h$, $\Delta_{\star}=\Delta /r_h^2$, $a_{\star} = a/r_h$, $\lambda_{\star}=\lambda r_h^2$ and $\Lambda_{\star}=\Lambda r_h^2$,
the radial equation takes the form \cite{Kodama:2009rq}: 
\beq
\frac{d^2\Phi}{dy_{\star}^2}+ Q(x)\Phi=0~,
\label{WKB}
\eeq
where
\beq
Q(x)=\left[\left(\omega_{\star}-\frac{a_{\star}m(1+a_{\star}^2) \biglb(1- {\lambda_{\star} } \bigrb)}{(x^2+a_{\star}^2)^2x^{n-1}} \right)^2-\frac{\Delta_{\star}(x)}{(x^2+a_{\star}^2)^2}U(x)\right]~,
\eeq
with
\beq
\Delta_{\star}(x)=( x^2 + a_{\star}^2) ( 1-{\lambda_{\star}^2} x^2) -  x^{1-n}(1+a_{\star}^2)(1-\lambda_{\star})~,
\eeq
and
\bea
U(x) &=& A_{ljm} -\frac{a_{\star}^2 m^2}{(x^2+a_{\star}^2)^2} \left((x^2+a_{\star}^2)(1+\lambda_{\star} a_{\star}^2)+ \frac{(1+a_{\star}^2)(1-\lambda_{\star})}{x^{n-1}} \right) + \frac{n(n+2)}{4}(1-\lambda_{\star} a_{\star}^2) -\frac{(n+2)(n+4)}{4}{\lambda_{\star}} x^2 \nn 
  &&  -\lambda_{\star} a_{\star}^2+\left(j+\frac{n}{2}\right)\left(j+\frac{n}{2}-1\right) \frac{a_{\star}^2}{x^2} +\frac{a_{\star}^2(1+\lambda_{\star} a_{\star}^2)}{x^2+a_{\star}^2} +\frac{\left((n+2)x^2+n a_{\star}^2\right)^2-8 a_{\star}^2 x^2}{4(x^2+a_{\star}^2)^2} \frac{(1+a_{\star}^2)(1-\lambda_{\star})}{x^{n-1}}~.
\eea
\noindent Since $\Delta\rightarrow 0$ at the horizons this form of the radial equation will be useful in determining the near and far field solutions in section \ref{sec:greybody}.


\section{Spheroidal Eigenvalues}\label{sec:eval}

\par It was found that the Continued Fraction Method (CFM) was the fastest way to generate the eigenvalues in this case, see reference \cite{Cho:2009wf} for other approaches. To do this we rewrite equation (\ref{angular}) in terms of the variable $x=\cos(2\theta)$, $c=a \omega$ and $\tilde\alpha= a^2\lambda$ \cite{Kodama:2009rq}: 
\bea
\label{angular-part}
(1-x^2) (2 + \tilde\alpha (1+x))S''(x)+\bigglb( n-1 -(n+3)x + {\tilde\alpha\over 2}(1+ x)(n+1 - (n+5)x) \biggrb)S'(x)&&\nonumber\\
+\Bigglb({A_{kjm}\over 2}+\frac{c^2 (x-1)}{2(2+\tilde\alpha (1+x))}+\frac{m^2 (1 + \tilde\alpha)}{x-1}-\frac{j (j + n-1)}{x+1}\Bigglb)S(x)&=&0 \,\,\, , 
\eea
and define $x=2z-1$, with the mode functions scaled by the characteristic exponents: 
\beq
Q(x)=  2^{|m|\over 2} (z-1)^{|m|\over 2}(2z)^{j\over2}\bigglb(z+{1\over \tilde\alpha}\biggrb)^{\pm {i c\over 2\sqrt{\tilde\alpha}} }y(z) \,\,\, .
\eeq
The angular mode equation can be written in the Heun form \cite{Suzuki:1998vy,Ron:1995}:
\beq
\bigglb[\frac{d^2}{dz^2}+\left(\frac{\gamma}{z}+\frac{\delta}{z-1}+\frac{\epsilon}{z+{1\over \tilde\alpha}}\right)\frac{d}{dz}+\frac{\alpha\beta z-q}{z(z-1)(z+{1\over\tilde\alpha})}\biggrb]y(z)=0 \,\,\, , 
\eeq
where (note that $\alpha$ should not to be confused with $\tilde\alpha$)
\bea
&& \alpha =\frac{1}{2} (j + |m| \pm i{c\over \sqrt{\tilde\alpha}} )\,\,\, , \qquad\qquad
   \beta = \frac{1}{2} (j + |m| + n+3 \pm i{c\over \sqrt{\tilde\alpha}}) \,\,\, ,\\
&& \gamma=\frac{1}{2} (2 j + n+1) \,\,\, , \qquad\qquad
   \delta =1+|m| \,\,\,, \qquad\qquad
   \epsilon = 1\pm i{c\over \sqrt{\tilde\alpha}} \,\,\, ,\\
&\mathrm{and} & q=-\frac{m^2}{4}+ \frac{1}{4}(j\pm  i{c\over \sqrt{\tilde\alpha}})(j+n+1\pm  i{c\over \sqrt{\tilde\alpha}})- {1\over 4 \tilde\alpha} \Big[ (j+|m|)(j+|m|+n+1) -A_{kjm}\Big] \,\,\, , 
\eea
with the constraint 
\beq
\alpha+\beta+1=\gamma+\delta+\epsilon \,\,\, .
\eeq
Note that these results are identical to the Kerr-AdS case considered by Kodama et al. \cite{Kodama:2009rq} by choosing $\tilde\alpha = -a^2/R^2$ with $c=a\omega$.

\par Since a three-term recurrence relation is guaranteed for any solution to Heun's differential equation \cite{Suzuki:1998vy,Ron:1995}, we can write:
\beq
\alpha_p c_{p+1}+\beta_p c_p+\gamma_p c_{p-1}=0 \,\,\, ,
\eeq
where for the Kerr-(A)dS case:
\bea
\alpha_p&=&-
\frac{(p + 1)(p + r - \alpha + 1)(p+ r -\beta + 1)(p + \delta)}
{(2p + r + 2)(2p + r + 1)}\,\,\,,\\
\beta_p&=&\frac{\epsilon p(p+r)(\gamma-\delta)+[p(p+r)+\alpha\beta][2p(p+r)+\gamma(r-1)]}{(2p + r + 1)(2p + r - 1)}-{1\over\alpha}p(p+r)-q\,\,\,,\\
\gamma_p&=&-
\frac{(p + \alpha - 1)(p + \beta - 1)(p + \gamma - 1)(p + r - 1)}
{(2p + r - 2)(2p + r - 1)}\,\,\,,
\eea
with
\beq
r  = j+|m|+\frac{n+1}{2}\,\,\,.
\eeq
Note that from the $c\to 0$ limit we find that $2k=\ell-(j+|m|)$ must be equal to an integer \cite{Cho:2009wf}. Once a 3-term recurrence relation is obtained the eigenvalue $A_{kjm}$ can be found (for a given $\omega$) by solving a continued fraction of the form \cite{Leaver:1985ax, Berti:2005gp}:
\beq
\beta_0-\frac{\alpha_0\gamma_1}{\beta_1-}\frac{\alpha_1\gamma_2}{\beta_2-}\frac{\alpha_2\gamma_3}{\beta_3-}\ldots=0\,\,\,.
\label{CFM}
\eeq

\par
\begin{figure}[th]
\scalebox{0.75}{\includegraphics{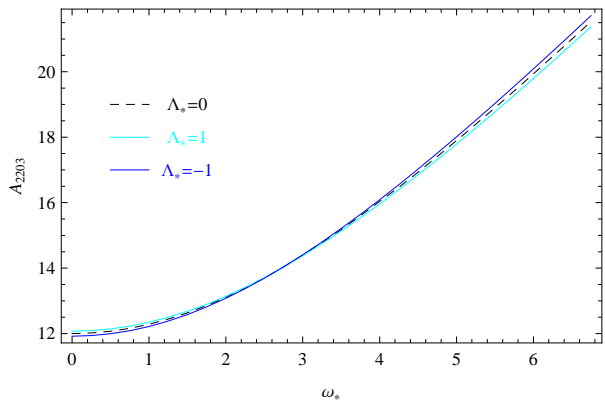}}
\caption{\it A typical eigenvalue plot for $(l,j,m,n)=(2,2,0,3)$, $a=1.2$ comparing the asymptotically flat ($\Lambda=0$), de Sitter ($\Lambda=1$) and anti de Sitter ($\Lambda=-1$) eigenvalues respectively.}
\label{fig:Evalue} 
\end{figure}

 It may be worth mentioning that in order to find the eigenvalues in the flat case, we could not simply take the limit $\lambda\rightarrow 0$ (using the Asymptotic Iteration Method (AIM) this limit is possible \cite{Cho:2009wf}). The CFM for the flat case was instead implemented using the input parameters described in reference \cite{Berti:2005gp}. In Fig. \ref{fig:Evalue} we have plotted the eigenvalues for several choices of the cosmological constant.

\section{Kerr-dS Greybody factors}\label{sec:greybody}

\subsection{Black Hole Horizon Limit}

\par In the near the horizon limit, $x\to 1$, the radial solution $\Phi$ has the following form (for {\it IN} modes \cite{Casals:2008pq,Kodama:2009rq}):
\bea
\Phi_{\rm NH} &=&A_{\rm in}^{(H)}  e^{-i\tilde\omega_{\star} x_\star}+A_{\rm out}^{(H)}  e^{-i\tilde\omega_{\star} x_\star}~,
\label{horiz}
\eea
where 
\beq
\tilde\omega_{\star}=\omega_{\star}-m\Omega_{\star}~, \qquad\qquad\qquad
\Omega_{\star} = {a_{\star}(1-\lambda_{\star})\over(1+a_{\star}^2)}~.
\eeq

\par Imposing that there are no outgoing modes at the black hole horizon implies the following boundary value problem for {\it IN} modes: $A^{(H)}_{\rm out}=0$, where without loss of generality we choose the normalisation of $A^{(H)}_{\rm in}$ such that we have the following initial value problem:
\bea
\Phi_{\rm NH}(x_0)&=& 1~, \nn
\Phi'_{\rm NH}(x_0)&=& -i\tilde\omega_{\star} {x_0^2 + a_{\star}^2 \over \Delta_{\star} (x_0)}~,
\eea
where $x_0= 1+ \epsilon$ with $\epsilon \sim 10^{-5}$. The greybody factor can be determined numerically by integrating the Schr\"odinger equation (\ref{WKB}) with the above horizon IVP matched onto the appropriate far-field form, see below. 


\subsection{Far Field \& Cosmological Horizon limit}

\par
\begin{figure}[th]
\scalebox{0.4}{\includegraphics{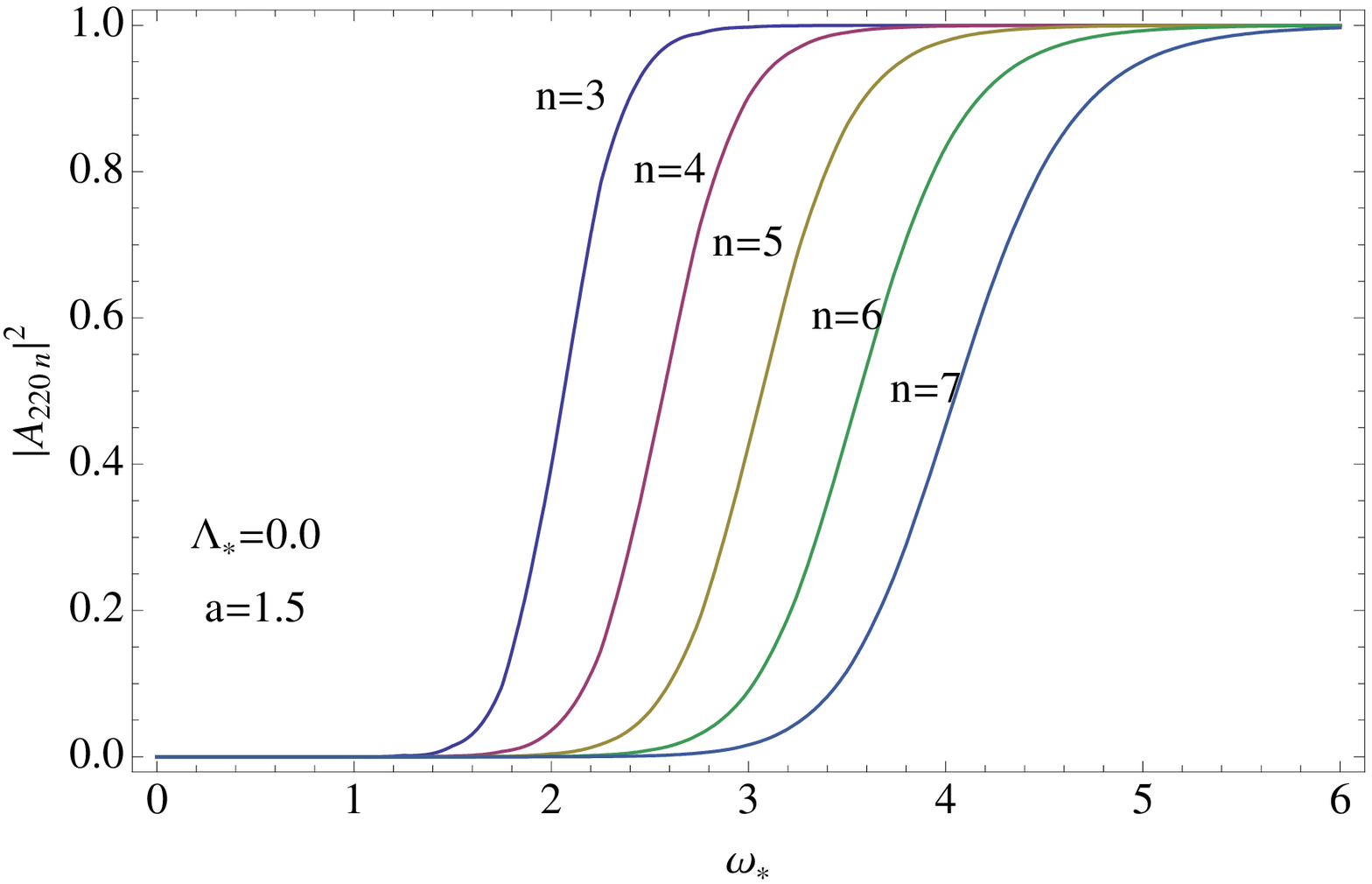}}
\hspace{0.5cm}
\scalebox{0.4}{\includegraphics{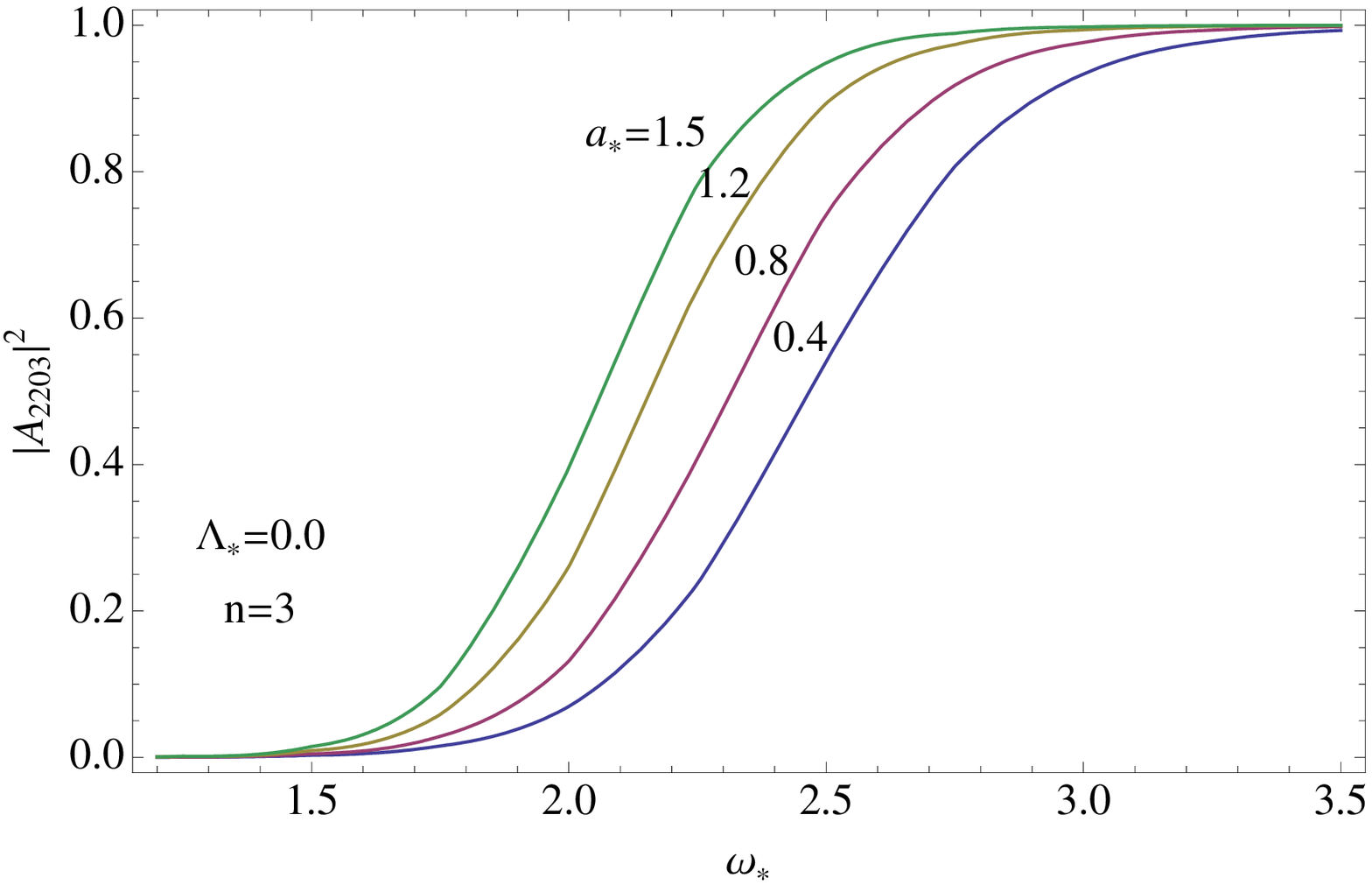}}
\scalebox{0.4}{\includegraphics{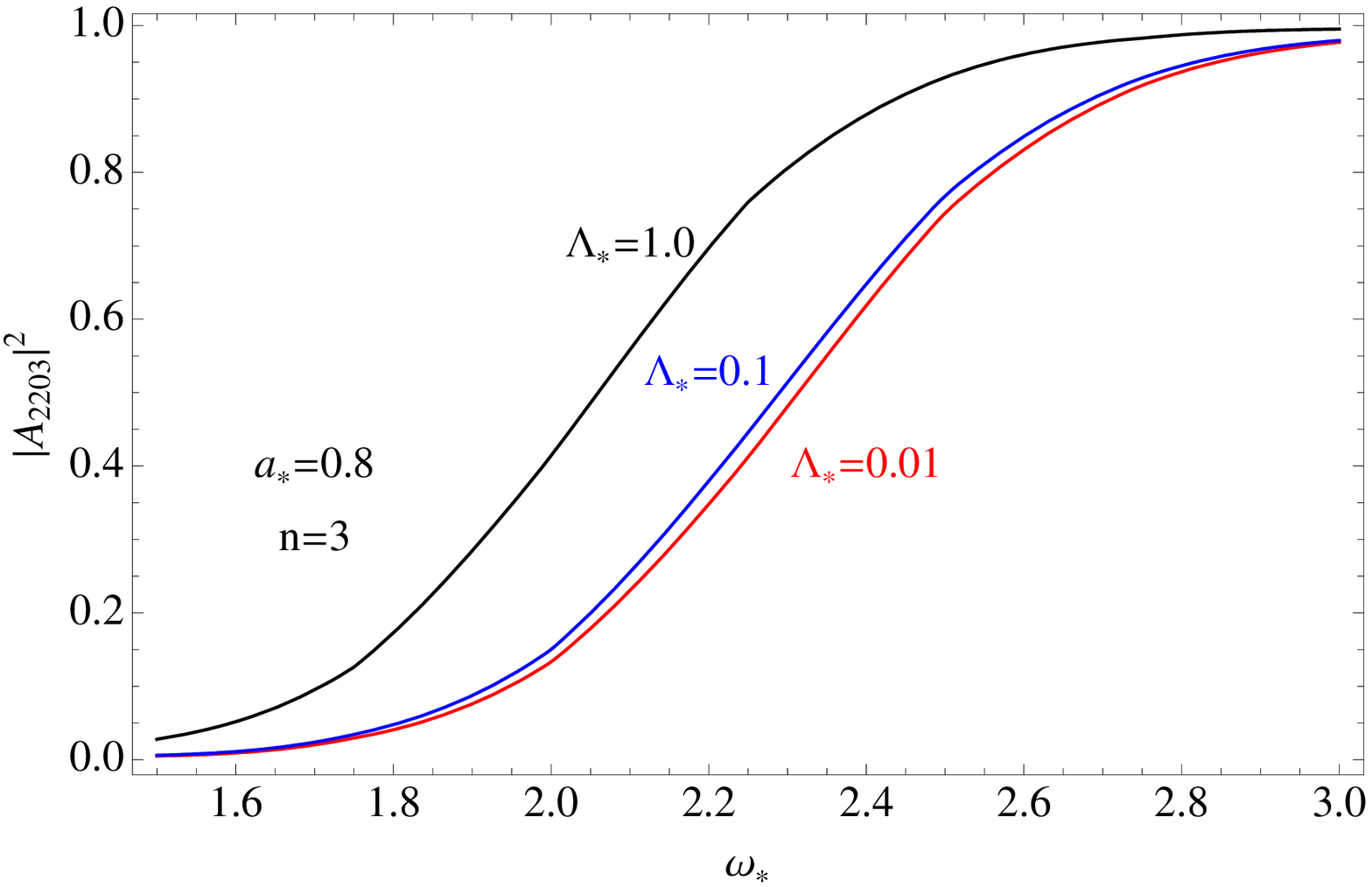}}
\hspace{0.5cm}
\scalebox{0.4}{\includegraphics{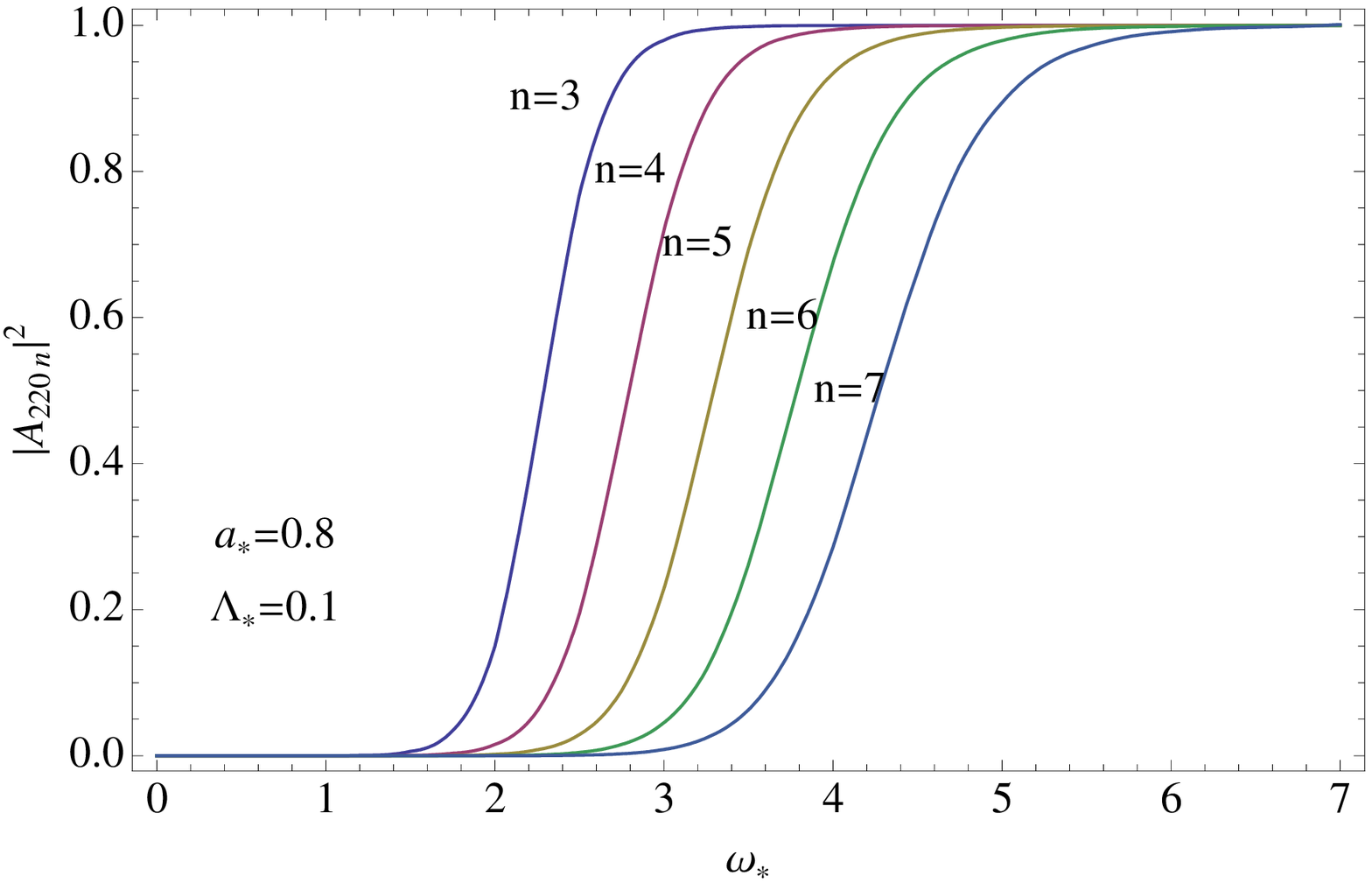}}
\caption{\it Various plots of absorption probabilities in the asymptotically flat case, where unless stated all plots are for $n=3$. Note that on the scale of these plots superradiance is too small to discern.}
\label{fig:Abs}
\end{figure}

 First consider the Kerr-AF  case, the solutions have a far field (FF) form at spatial infinity, where $x_{\star} \to x$ (for $x\to\infty$):
\beq
\Phi_{\rm FF}\approx x^{-\tfrac{n+1}{2}}\left(A_{\rm in}^{(\infty)} e^{-i\omega_\star x} +A_{\rm out}^{(\infty)}  {e^{i\omega_\star x} }\right)~.
\label{farf}
\eeq
The NH solution can then be matched onto the FF equation (\ref{farf}) \cite{Ida:2005ax}, where  the reflection coefficient is then defined as the ratio $|{\cal R}_{ljm}|^2={|A_{\rm out}^{(\infty)} |^2/ |A_{\rm in}^{(\infty)} |^2}$, and the relationship between the absorption and reflection coefficient is:
\beq
|{\cal A}_{ljmn}|^2=1-|{\cal R}_{ljmn}|^2=1-\left|{A_{\rm out}^{(\infty)} \over A_{\rm in}^{(\infty)} }\right|^2~.
\label{absref}
\eeq 
Some typical examples of the absorption probability as a function of $\omega_{\star}$ (=$\omega r_h^2$) in the asymptotically flat limit are shown in the top two panels of Fig. \ref{fig:Abs}. 

\par As opposed to the Kerr-AF case, the Kerr-de Sitter case is a little more subtle \cite{Kanti:2005ja}. In this case we should instead assume the form of both ingoing and outgoing waves at the cosmological horizon:
\beq
\Phi_{\rm c}\approx A_{\rm in}^{(c)} {e^{-i\omega_{\star} y_\star} }+A_{\rm out}^{(c)}  {e^{i\omega_{\star} y_\star} }~,
\label{coshor}
\eeq
with the tortoise coordinate given in equation (\ref{tort}) in dimensionless variables and with no tilde on the omega. The above expression works well for the case of a small cosmological constant $\lambda_{\star} a_{\star}^2 \ll 1$ (that is where $r_c\gg r_h$), where
\beq
|{\cal A}_{ljmn}|^2=1-|{\cal R}_{ljmn}|^2=1-\left|{A_{\rm out}^{(c)} \over A_{\rm in}^{(c)} }\right|^2~.
\label{absdS}
\eeq 
Note for the choices of parameters presented in this paper we found that a coordinate transformation along the lines of that done in the Schwarzchild de Sitter case in reference \cite{Kanti:2005ja}, which moves the cosmological horizon essentially to infinity, was not required. Results for the Kerr-dS case are given in the bottom two panels of Fig. \ref{fig:Abs}, where we see that a larger cosmological constant has a similar effect to increasing the rotation, $a$, or lowering the dimensionality, $n$; that is, larger $\Lambda$ leads to lower energies at which there is full transmission of scattered waves.

\section{Superradiance}

\par
\begin{figure}[th]
\scalebox{0.55}{\includegraphics{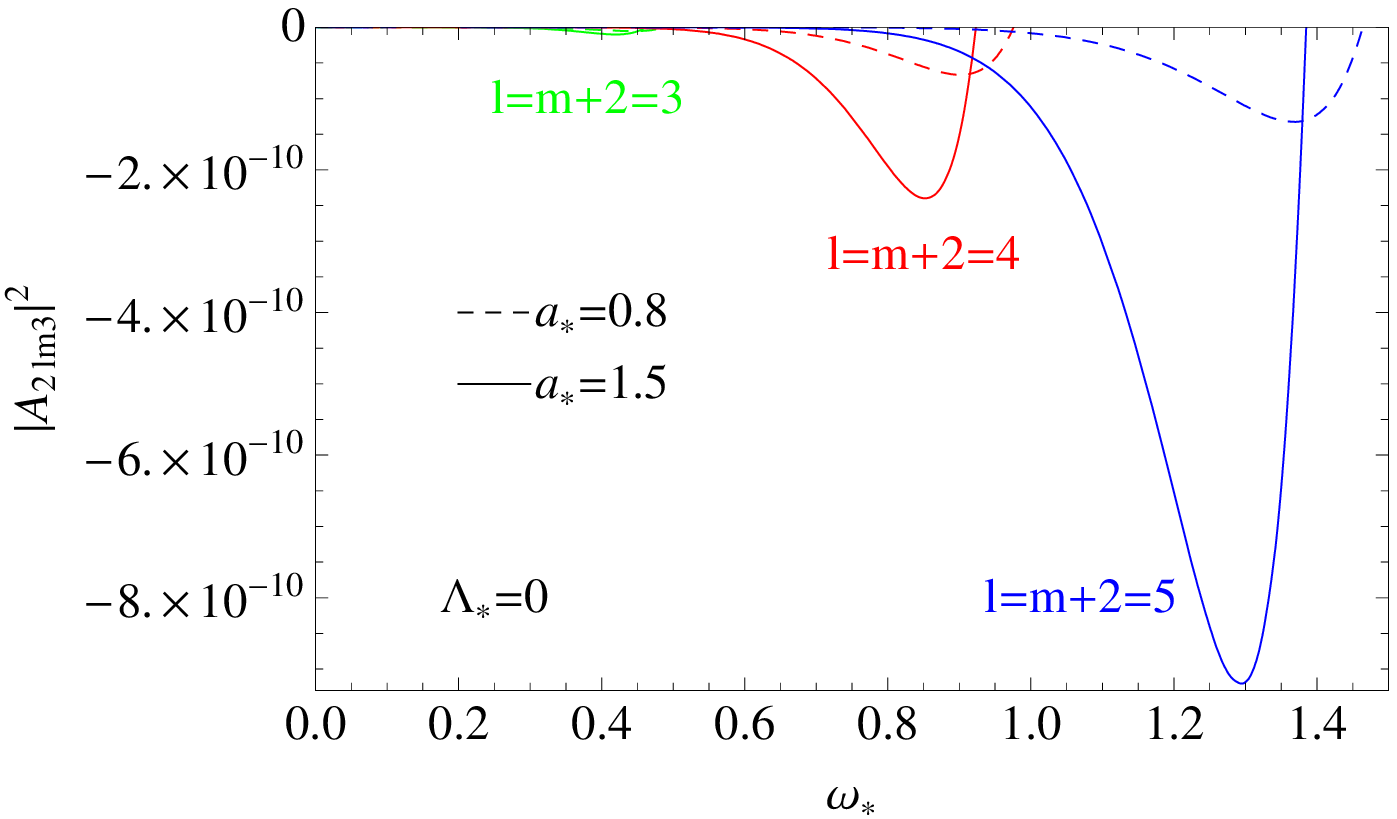}}{a}
\hspace{0.55cm}
\scalebox{0.55}{\includegraphics{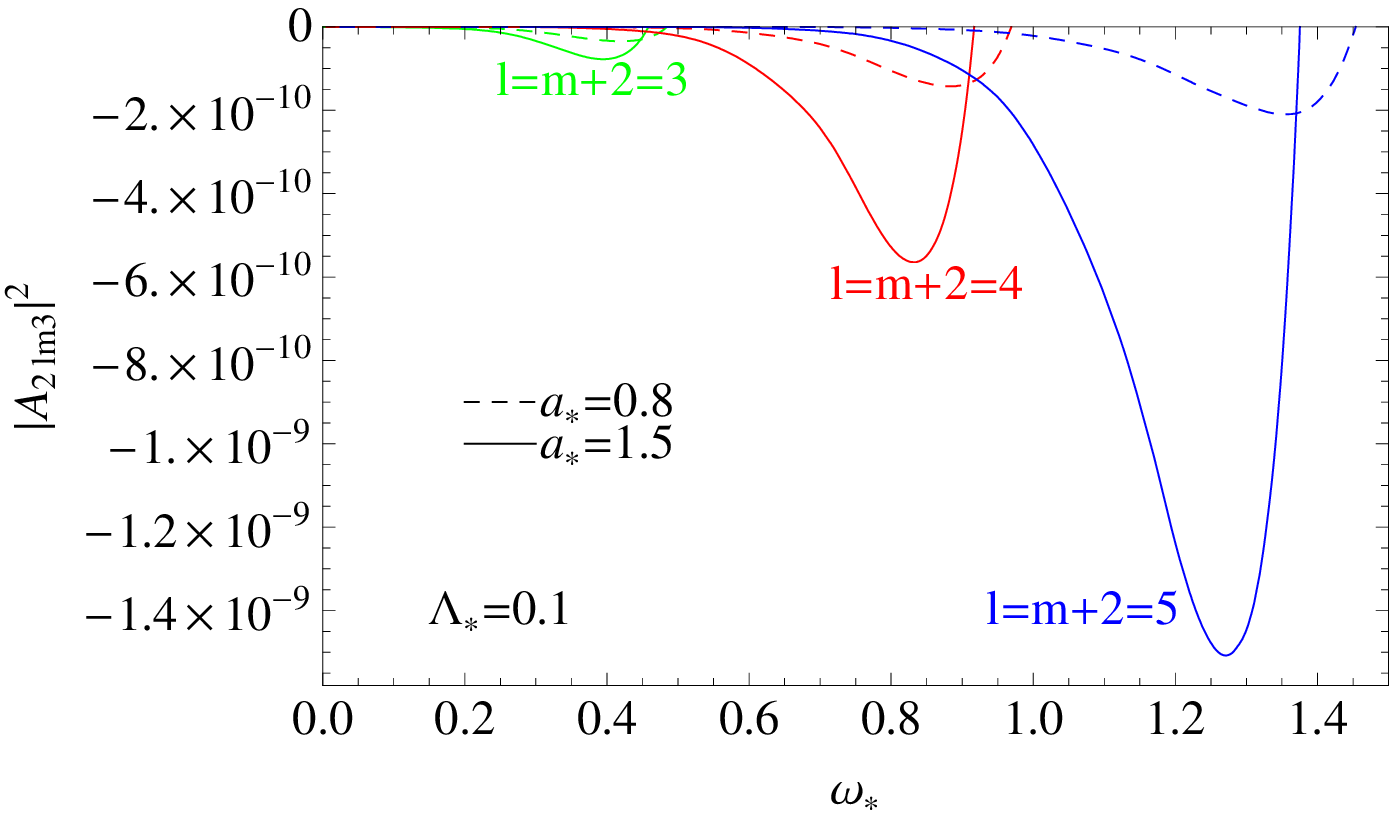}}{b}
\caption{\it Absorption probability plots in the superradiance regime for the asymptotically flat case (a) and for the de Sitter case (b) with cosmological constant ($\Lambda r_h^2=0.1$).}
\label{fig:AbsSR}
\end{figure}

 Employing the constancy of the Wronskian \cite{DeWitt:1975ys,Ottewill:2000qh}: $\Phi_1 d \Phi_2/dy -\Phi_2 d \Phi_1/dy$, which is valid for any solution $\Phi$ and its complex conjugate $\Phi^*$ (for real eigenvalues) allows one to find the following relation:
\beq
1-\left|{A_{\rm out}^{(\infty)} \over A_{\rm in}^{(\infty)} }\right|^2 = {\tilde \omega \over \omega} \left|{ A_{\rm in}^{(H)} \over A_{\rm in}^{(\infty)} }\right|^2~.
\label{SRWronsk}
\eeq
As is well known, solutions with charge or rotation undergo super-radiance \cite{DeWitt:1975ys}, where the condition for super-radiance to occur is: 
\beq
{\tilde\omega\over\omega}=\frac{\omega-m \Omega }{\omega}=1-{m a_{\star}(1-\lambda_{\star})\over  \omega_{\star}(1+a_{\star}^2)} <0 ~.
\label{suprad}
\eeq
In this case the absorption probability becomes negative, as can be seen from equation (\ref{SRWronsk}). Some plots of the superradiance regime are shown for both the asymptotically flat and de Sitter cases in Fig. \ref{fig:AbsSR}. Note though that from the $1 - \lambda_*$ factor in equation (\ref{suprad}) the possible size of $\lambda$ is limited.

\par
\begin{figure}[th]
\scalebox{0.525}{\includegraphics{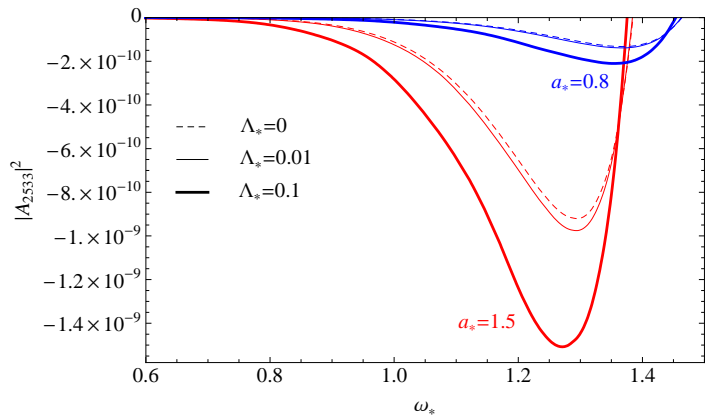}}
\caption{\it Superradiance plots for various choices of the cosmological constant ($\Lambda r_h^2$), for the rotations $a_{\star}=0.8$ (blue) and $a_{\star}=1.5$ (red).}
\label{fig:SRCompare}
\end{figure}

 An interesting feature of black holes in Kerr-dS spacetimes is that the superradiance effect is enhanced by the strength of the cosmological constant, this can be seen from Fig. \ref{fig:SRCompare}. 

\section{Surface Gravity Normalization}

\par Before progressing further, there are some subtle issues concerning the correct normalisation of the Hawking temperature in the emission formula for asymptotically de Sitter spacetimes \cite{Bousso:1996au, Kanti:2005ja}, where the definition of the surface gravity requires an extra normalisation.  This Bousso-Hawking temperature corresponds to the surface gravity as measured by an observer  where the black hole attraction and cosmological expansion cancels out; however, whether or not one should use this formula requires some careful consideration, see reference \cite{Myung:2007my,Wu:2008rb} and references therein. Indeed for the Kerr-dS case the generalisation of the Bousso-Hawking temperature appears to depend on the {\it latitude} angle $\theta$.

\par From equations (\ref{killsurf}, \ref{Lsq}) it is straightforward to show that
\beq
\kappa^2_{h,c}~= ~\stackrel{\lim}{{}_{r \to r_{h,c}} } \gamma^2 (\nabla_a L) (\nabla^a L)~,
\label{eqn:Surf}
\eeq
where for the Kerr-AF case the normalisation is chosen such that $\gamma_t^2 K^a_t K_{a,t}=-1$ or $\gamma^2=1$. However, it is important to note that for the Schwarzschild-dS case it is usually argued \cite{Bousso:1996au, Kanti:2005ja} that we cannot assume that $K^2 = -1$ (or $\gamma=1$) at spatial infinity ($r\to \infty$), because of the cosmological horizon at $r_c$. Instead we should find the radius $r_g$, for which the orbit of the Killing vector coincides with a geodesic going through  $r_g$ at constant angular $\phi$ variables (where the cosmological expansion and black hole attraction balances out). 

\par As we mentioned earlier,  for the Kerr-dS case it appears that  the value of $r_g$ depends on the latitude angle $\theta$ and suggests that the standard Hawking emission formula (integrated over angular variables) would have to be reconsidered. Interestingly, for the case of a higher dimensional simply rotating black hole (created on the brane) the latitude angle would be $\theta=\pi/2$ and so there appears to be no ambiguity with the choice of $\theta$. However, to avoid these subtleties we will make some simplifying assumptions, which leads to a form very similar to the Bousso-Hawking normalisation, for the Schwarschild-dS case. 

\par In the case where $a\lessapprox 1$, the line element in equation (\ref{metric}) can be approximated by something that looks like the standard Schwarzschild-dS metric:
\beq
ds^2 \approx -h(r,\theta,a) dt^2 + {1\over h(r,\theta,a)} dr^2 + r^2 d\Omega_{n+2}~,
\eeq
except where $h$ is a function of $\theta$ and $a$:
\beq
h(r,\theta,a) \equiv {\Delta_r \over \rho^2}= 1 - {2M \over (r^2+a^2\cos^2\theta ) r^{n-1}}
-{2 \Lambda  r^2\over (n+2)(n+3)} {r^2+a^2 \over (r^2+a^2\cos^2\theta )}=0~.\eeq 
In the case of  a case of small rotations and $\theta \sim \pi/2$ the form of $h(r,\theta,a)$ simplifies to
\beq
h(r,a) \approx 1 - {2M \over  r^{n+1}}
-{2 \Lambda  r^2\over (n+2)(n+3)} \Biglb(1+{a^2\over r^2}\Bigrb)=0~,
\eeq 
where we shall assume that the horizon, $r_h$, can still be expressed in terms of the Kerr-dS mass, as in equation (\ref{eqn:M}): $2M = r_h^{n-1} (r_h^2+a^2) (1- \lambda r_{h}^2 )$. Following the Bousso-Hawking argument \cite{Bousso:1996au,Kanti:2005ja} the actual Hawking temperature will then become:
\beq
T_{bh} ={T_h\over \sqrt{h(r_g,a)} } ={\kappa_h\over 2\pi \sqrt{h(r_g,a)} }~,
\label{BoussoT}
\eeq 
where $r_g$ is the solution of $h'(r,a)=0$. Quite interestingly the solution of $h'(r,a)=0$ is independent of $a$ in this approximation (solutions to $h(r,a)=0$ do depend on $a$):
\beq
r_g= \left(\frac{M (n+1) (n+2) (n+3)}{2 \Lambda}\right)^{\frac{1}{n+3}}~.
\eeq
This can be seen from the fact that for a given $r_g$ the term $\Lambda (1 + a^2/r_g^2)$ looks like a shifted cosmological constant, $\tilde\Lambda$. In the next section we shall evaluate emissions for the Hawking temperature, $T_h$, and the Bousso-Hawking-like one, $T_{bh}$.

\begin{figure}[th]
\scalebox{0.45}{\includegraphics{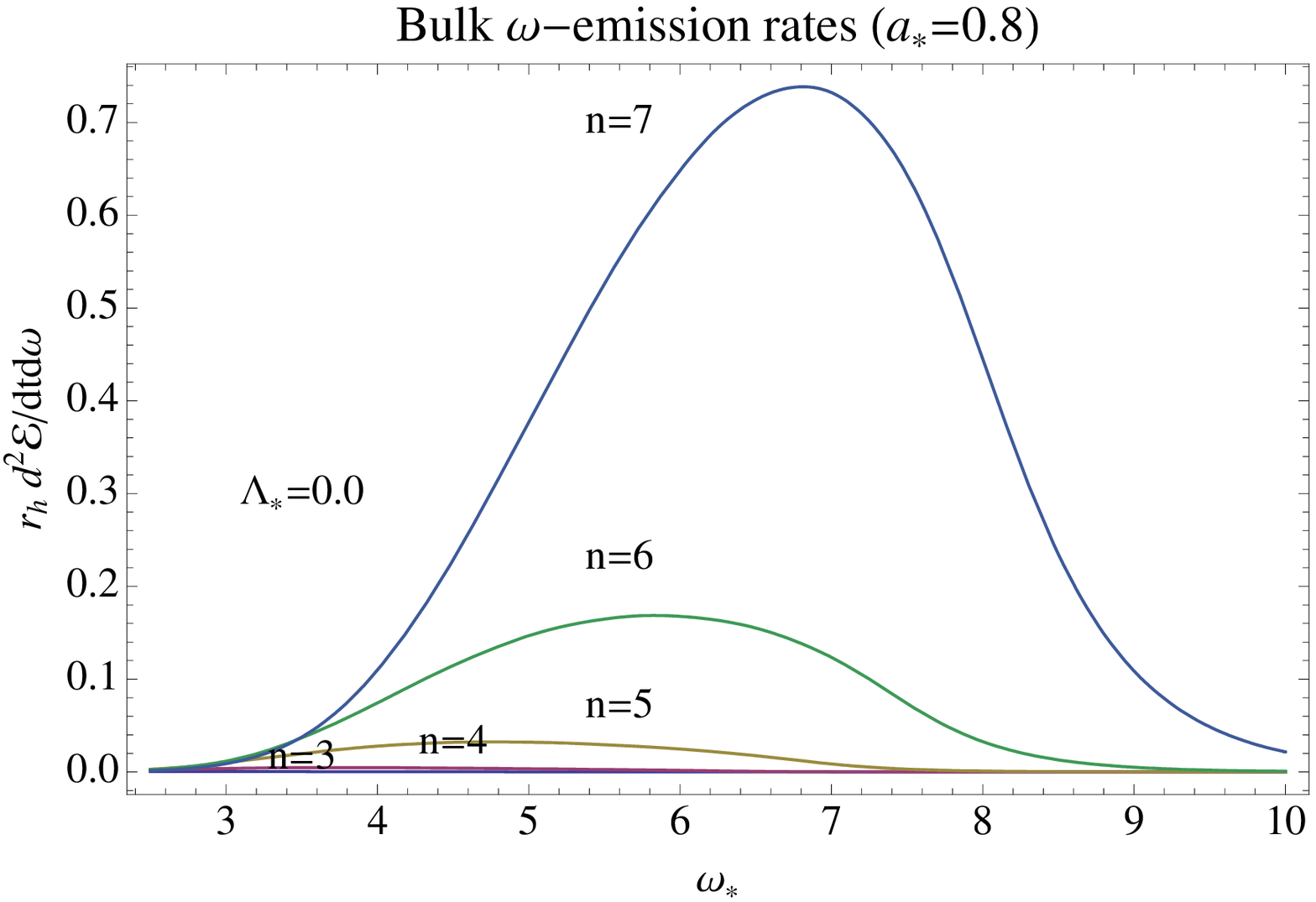}}
\hspace{0.5cm}
\scalebox{0.45}{\includegraphics{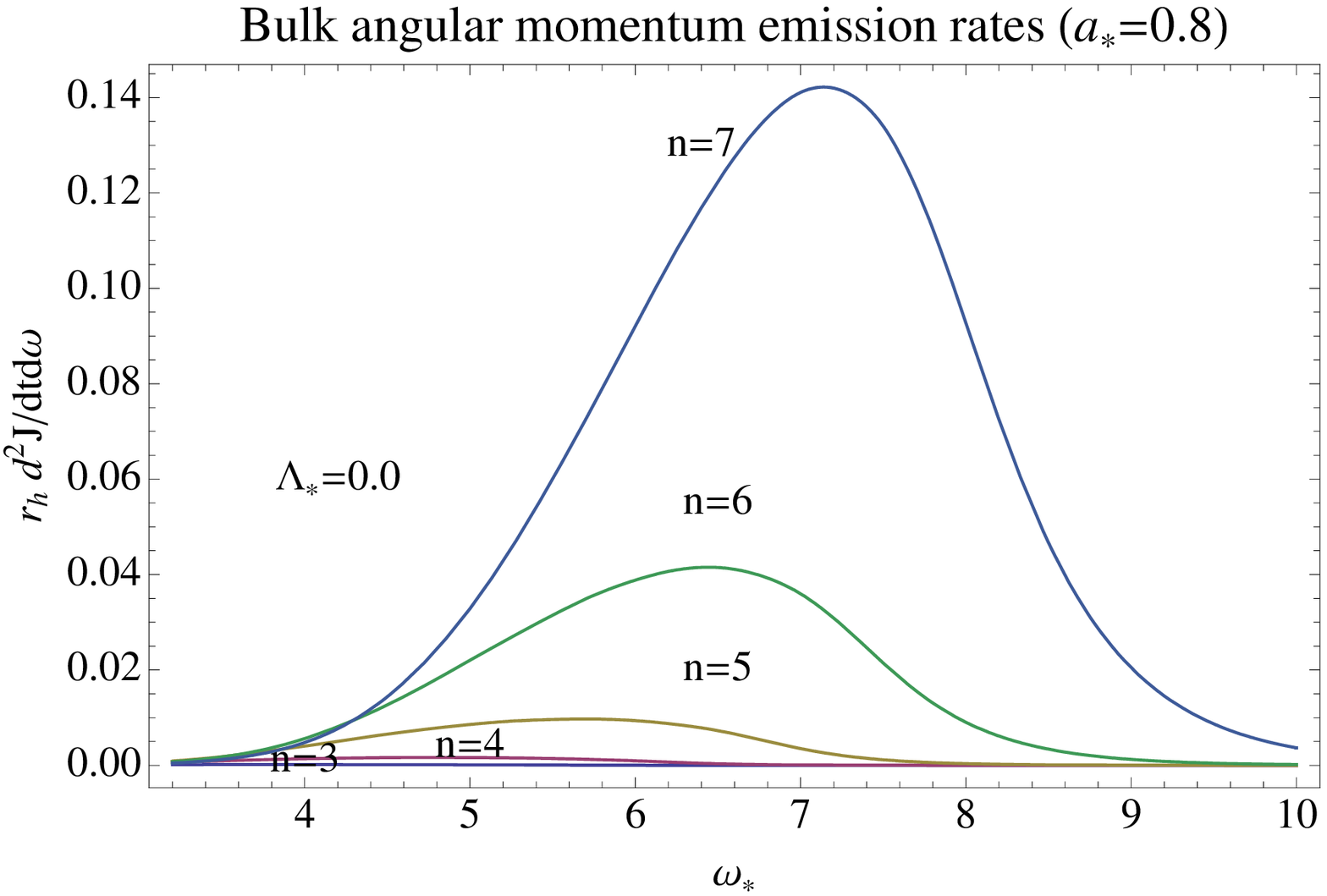}}
\caption{\it Energy and angular momentum emissions in asymptotically flat space ($\Lambda=0$) for different dimensions $n+4$ with $a/r_h=0.8$.}
\label{EmissAF1}
\end{figure}

\begin{figure}[th]
\scalebox{0.45}{\includegraphics{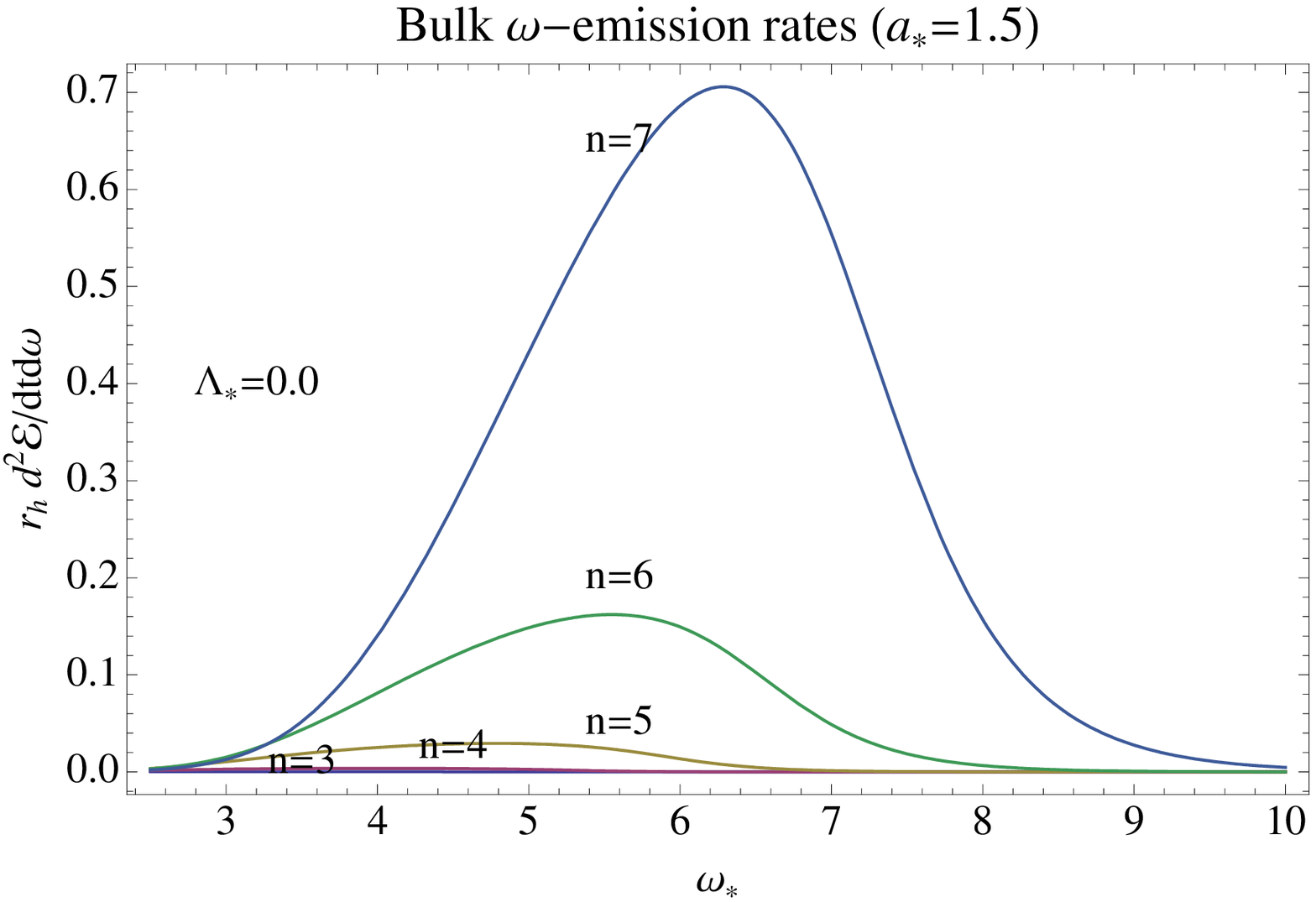}}
\hspace{0.5cm}
\scalebox{0.45}{\includegraphics{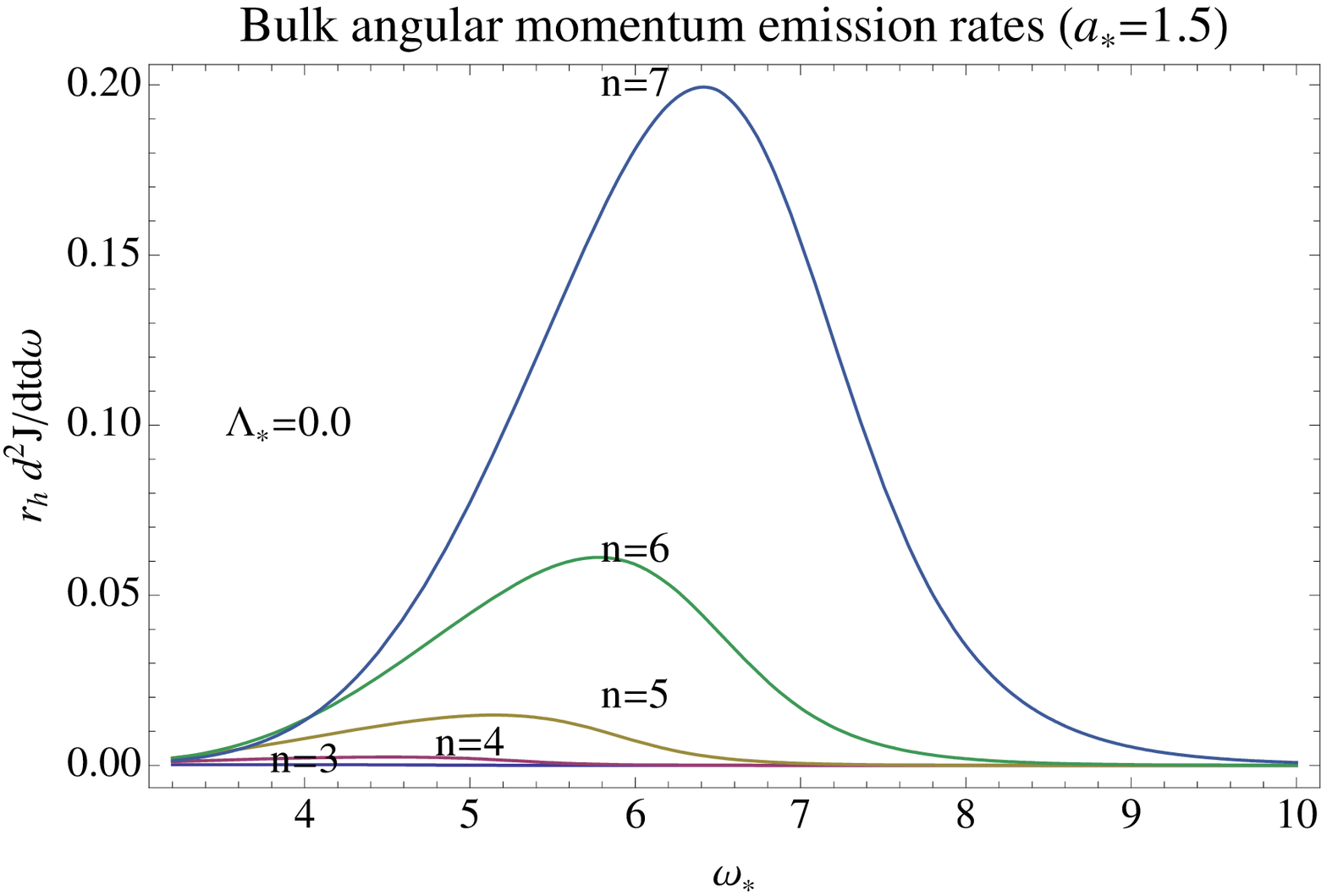}}
\caption{\it Energy and angular momentum emissions in asymptotically flat space ($\Lambda=0$) for different dimensions $n+4$ with $a/r_h=1.5$.}
\label{EmissAF2}
\end{figure}

\begin{figure}[ht]
\scalebox{0.35}{\includegraphics{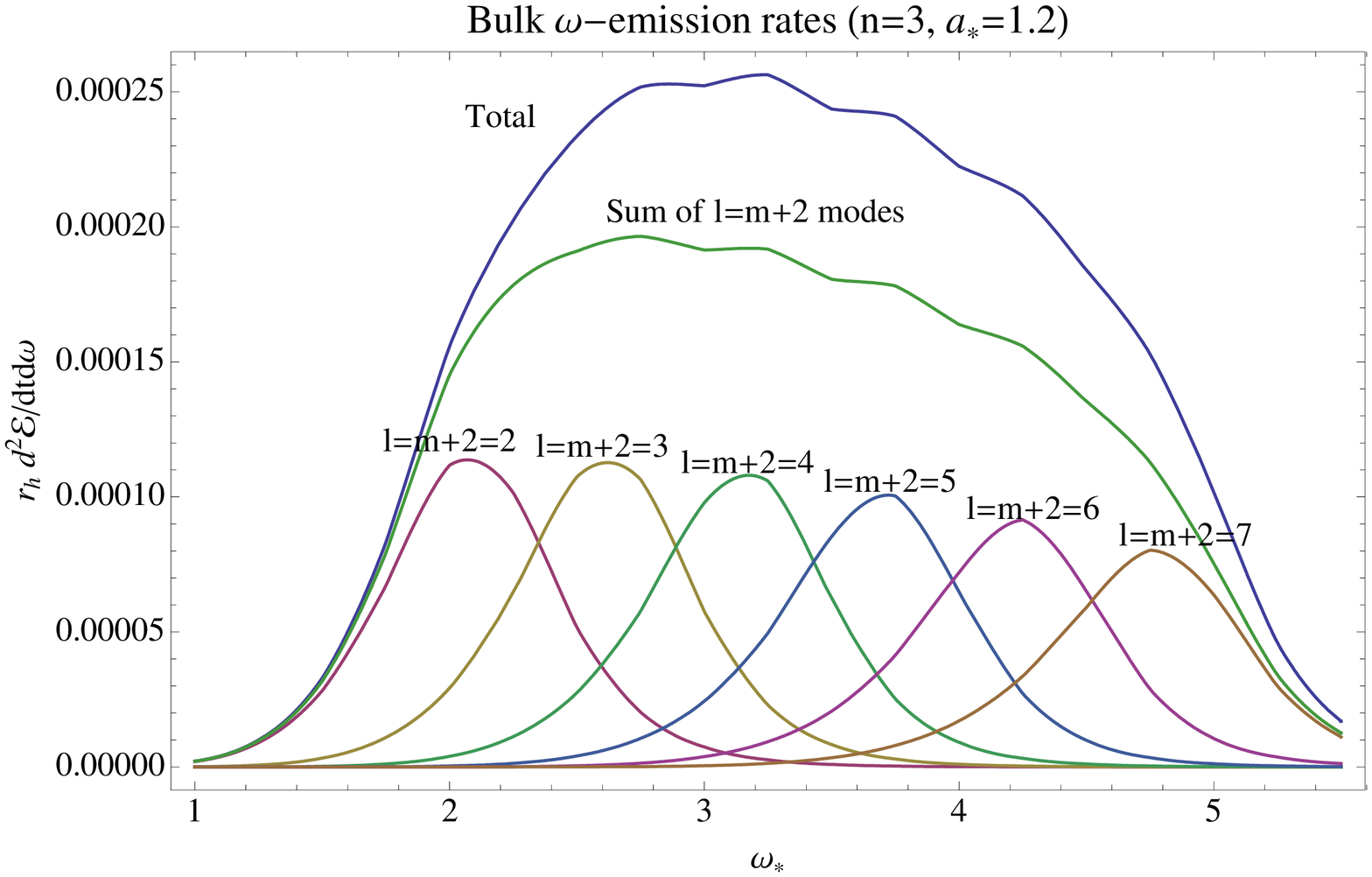}}
\hspace{0.5cm}
\scalebox{0.35}{\includegraphics{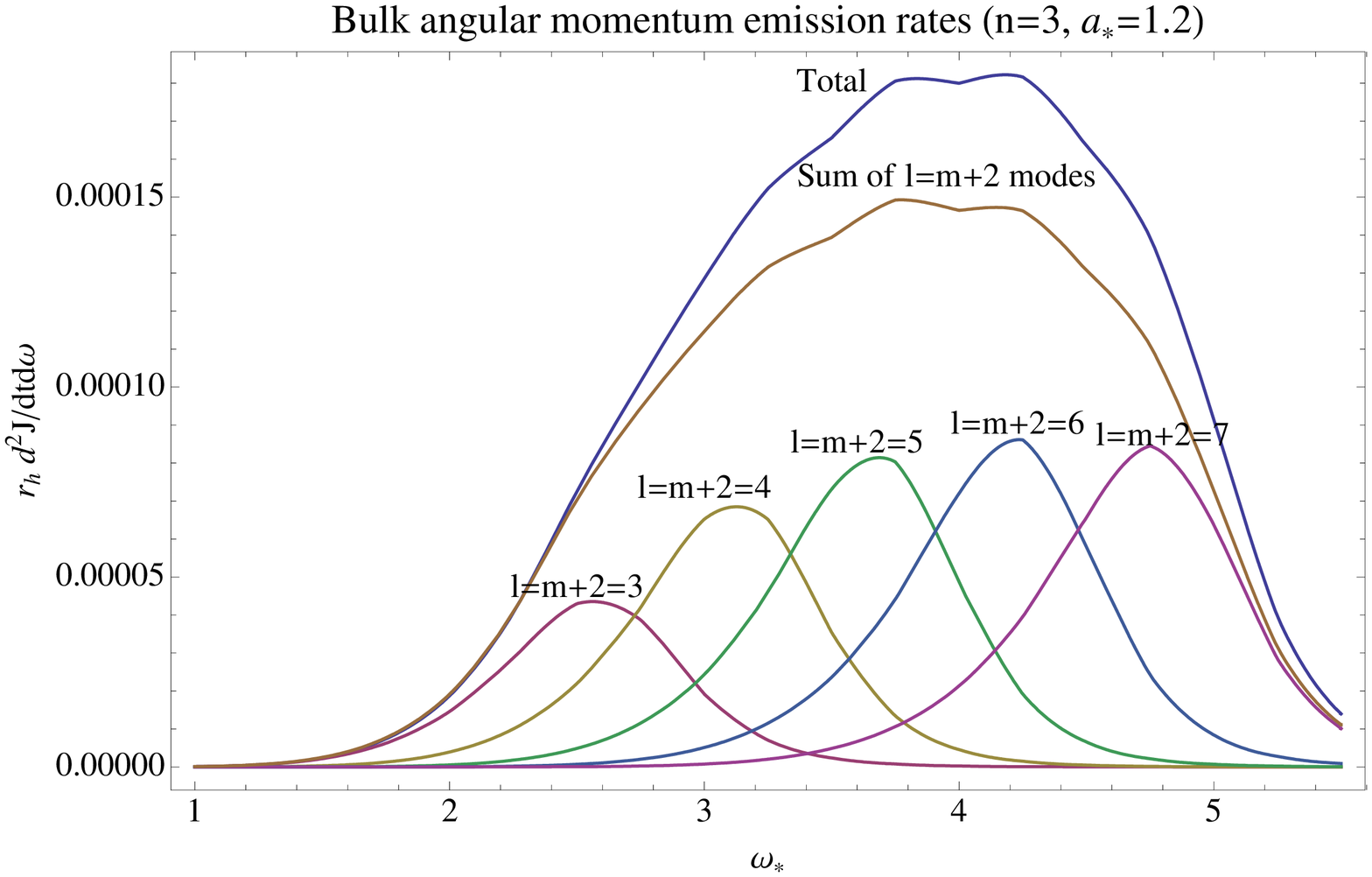}}
\caption{\it Contribution to energy and angular momentum emissions from the dominant $l=m+ 2=3, \dots$ modes for $n=3$ and $a_\star=1.2$ (with $\Lambda=0$).}
\label{EmissCont}
\end{figure}


\section{Energy and Angular Emission Rates}

\par Following reference \cite{Casals:2008pq} the quantization of the mode functions used in the stress energy tensor with appropriate choice of {\it UP} and {\it IN} modes, after integrating over the angular variables, using the standard normalizations, leads to:
\beq
 { d^2E\over dt \,d\omega} = {1 \over 2\pi} \sum_{j,l,m} {\omega \over \exp[{\tilde \omega\over T_h}]-1} D_j^{T} \left(1-|{\cal R}_{ljm}|^2\right) ~
\eeq
and
\beq
 { d^2J\over dt \,d\omega} = {1 \over 2\pi} \sum_{j,l,m} {m \over \exp[{\tilde \omega\over T_h}]-1} D_j^{T}
\left(1-|{\cal R}_{ljm}|^2\right) ~,
\eeq
where the Hawking temperature is related to the surface gravity
by \cite{Creek:2007pw}:
\beq
T_h ={ \kappa_h\over 2\pi}~.
\label{HawkT}
\eeq 
Note that as well as the Hawking temperature, $T_h$, there is also a Bousso-Hawking-like temperature, $T_{bh}$, which might be the more appropriate choice for the Kerr-dS case. Following arguments similar to that in reference \cite{Creek:2007pw} it is possible to show that as $a\to 0$, the Schwarzschild emission formulae for tensor modes are reproduced.

\par By direct comparison with equation (\ref{eqn:Surf}) the surface gravity, $\kappa_h$, at the horizon, for a simply rotating Kerr-dS black hole can be expressed in the simpler notation \cite{Myers:1986un,Hawking:1998kw}:
\beq
\kappa_h= ~ \lim_{ r\to r_{h,c} } \Biglb[ {\Delta_r \over 2 (r_h^2+a^2)(r-r_h)} \Bigrb] =  \Biglb[ {\Delta'_r(r_h) \over 2 (r_h^2+a^2) }\Bigrb]~.
\eeq
This implies that for the Kerr-dS case we have:
\beq
\kappa_h ={\left((n-1)-(n+1) \lambda_{\star}\right) a_{\star}^2+ (n+1)-(n+3) \lambda_{\star} \over 2 r_h (1+a_{\star}^2)} ~.
\eeq
A useful check is that  the asymptotically flat limit, $\lambda_{\star} a_{\star}^2\to 0$, agrees with the Myers and Perry result \cite{Myers:1986un,  Creek:2007pw, Casals:2008pq}:
\beq 
\kappa_h={(n+1)+(n-1)a_{\star}^2\over 2r_h(1+a_{\star}^2)}~,
\eeq
and that it agrees with higher-dimensional Schwarzschild de Sitter (SdS) limit, $a\to 0$, \cite{Kanti:2005ja}: 
\beq
\kappa_h= \frac{(n+1)-(n+3) \lambda  r_h^2}{2 r_h}~.
\eeq

\par For the Kerr-AF case, which is the main focus of this work, some of the results can be seen in Figs. \ref{EmissAF1} and \ref{EmissAF2}. The results are consistent with those of other works \cite{Casals:2008pq}, where they considered the bulk emission of scalar spin-$0$ fields on the Kerr-AF background. First of all, we see that both the bulk emission and angular momentum increases with increasing dimension. Second of all, the bulk emission decreases with increasing $a$, although for a given value of $a$, we expect the energy to start increasing. For bulk angular momentum, increasing $a$ leads to larger values as can be seen. Note that as was discussed in references \cite{Ida:2005ax, Harris:2005jx} the value of $a_*$ is bounded (from estimates arising from brane localised accelerator collisions). As such, we have not gone higher than $a_*=1.5$ in our simulations.

\par An important difference is that because the modes start from $j=2,3,4,\dots$ the spectrum is shifted to the right (larger $j$ corresponds to larger scattering energies $\omega$). For spin-$0$ fields the sums start from $j=0$, which implies lower energy emissions. The lack of $j=0,1$ modes has another effect, which can be seen from Fig. \ref{EmissCont}. The different mode contributions are shown here and is similar to the cases for scalar-fields on Kerr-AF \cite{Casals:2008pq} and for the tense brane variant \cite{Kobayashi:2007zi}. However, in general, the actual energy and angular momentum emissions are very much smaller and this stems from the fact that $j=0,1$ lead to large emissions. We will comment on the Kerr-dS results in the concluding remarks.

\begin{figure}[ht]
\scalebox{0.45}{\includegraphics{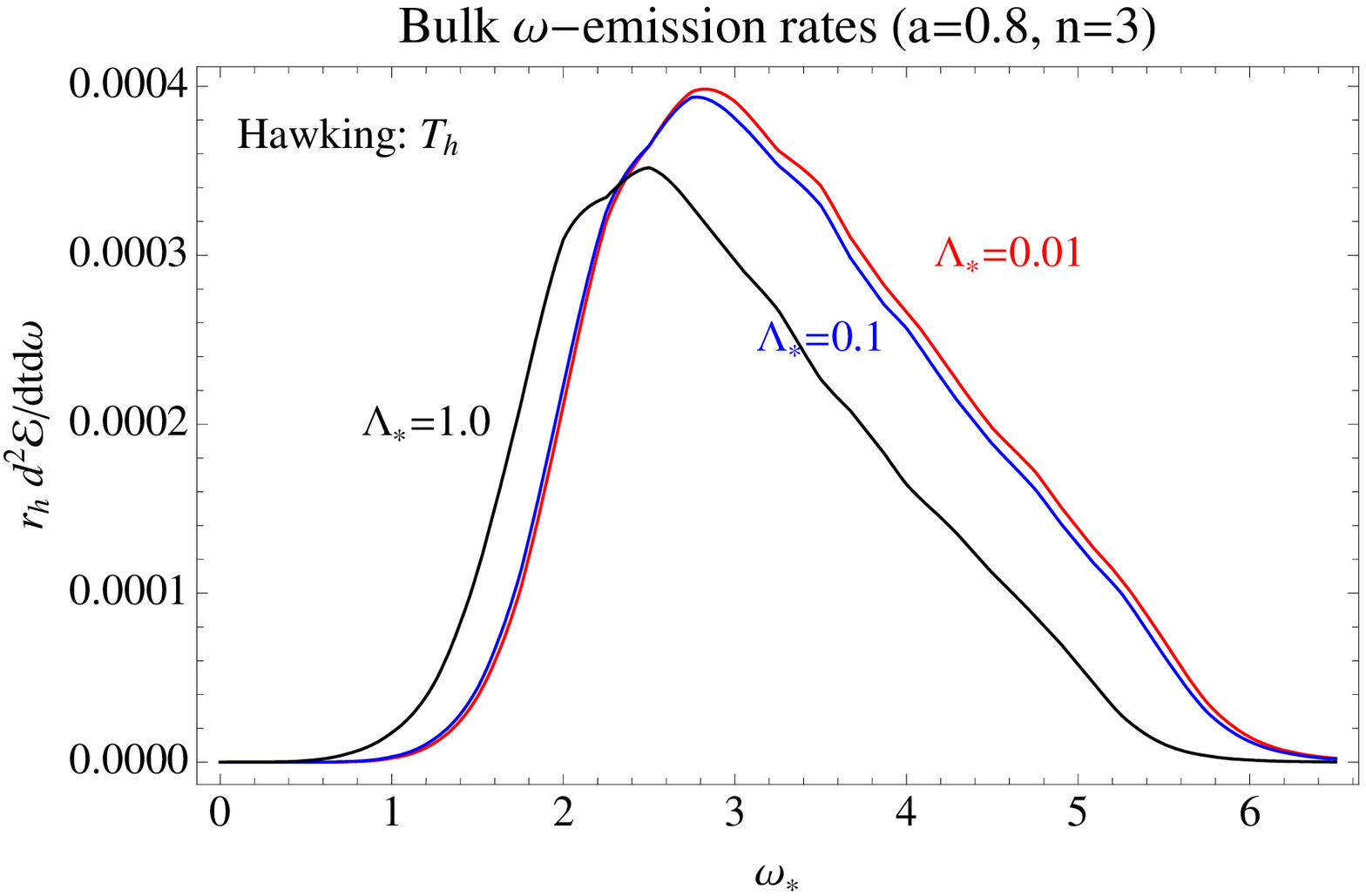}}
\hspace{0.5cm}
\scalebox{0.45}{\includegraphics{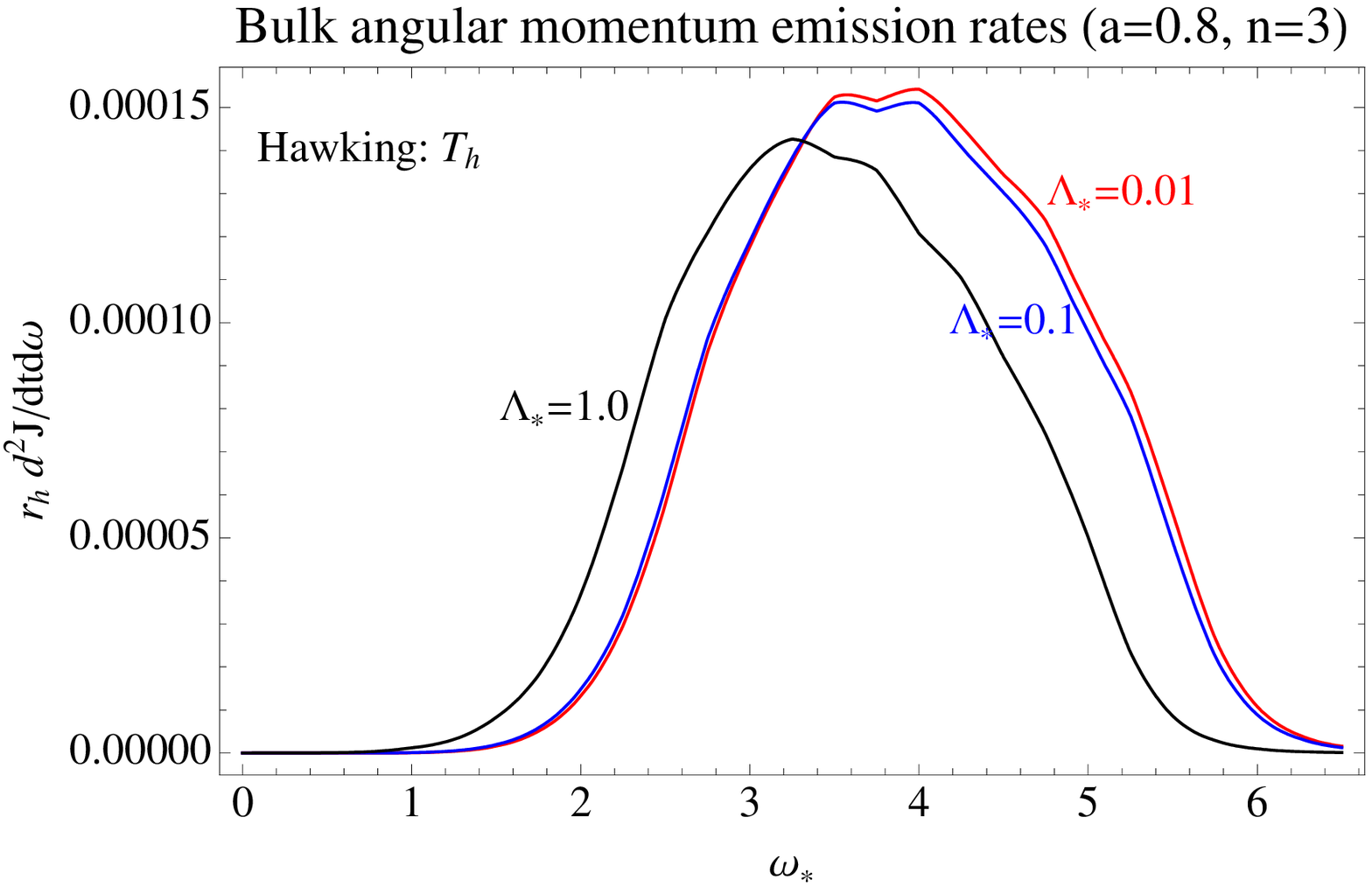}}
\scalebox{0.45}{\includegraphics{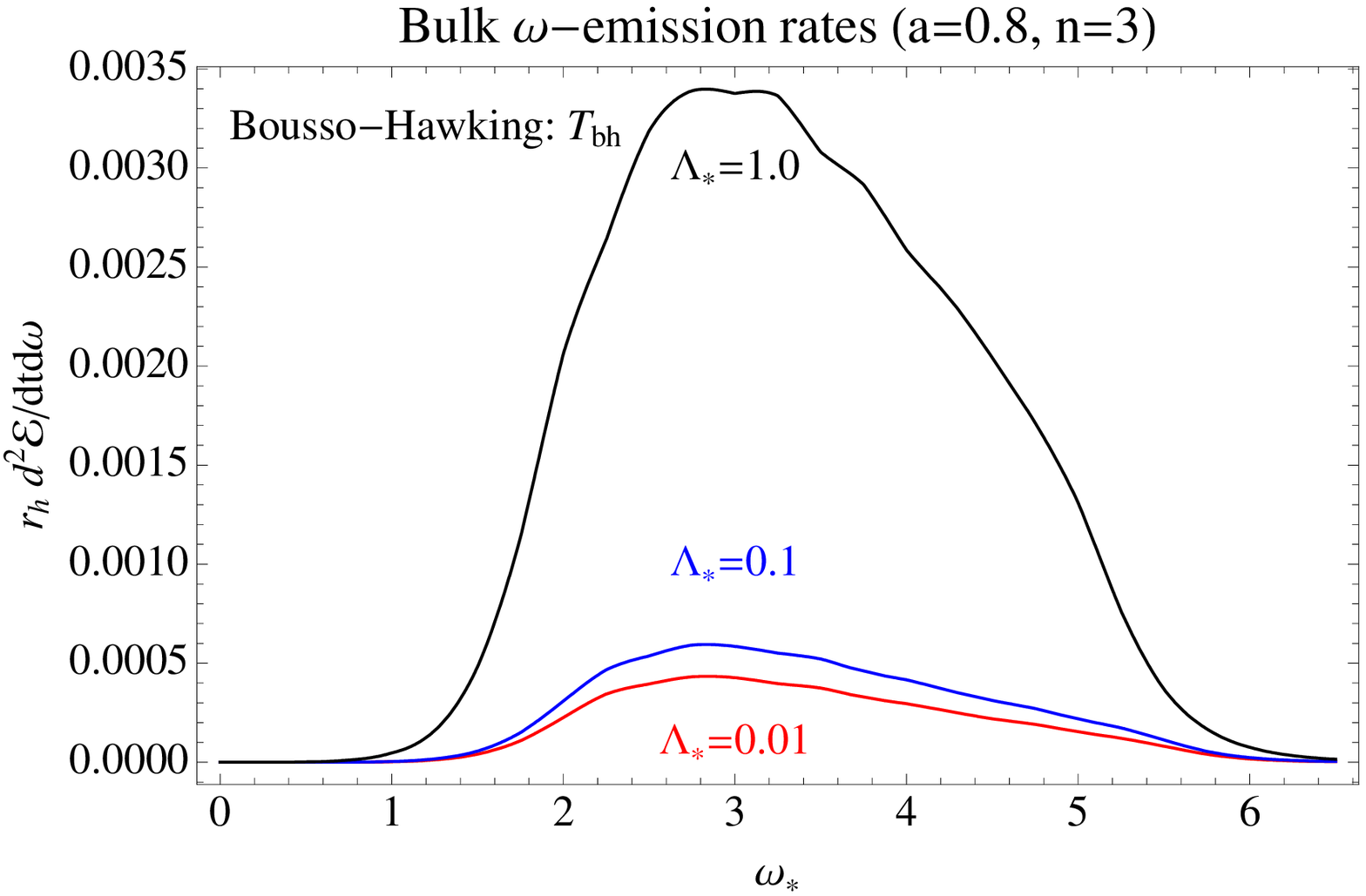}}
\hspace{0.5cm}
\scalebox{0.45}{\includegraphics{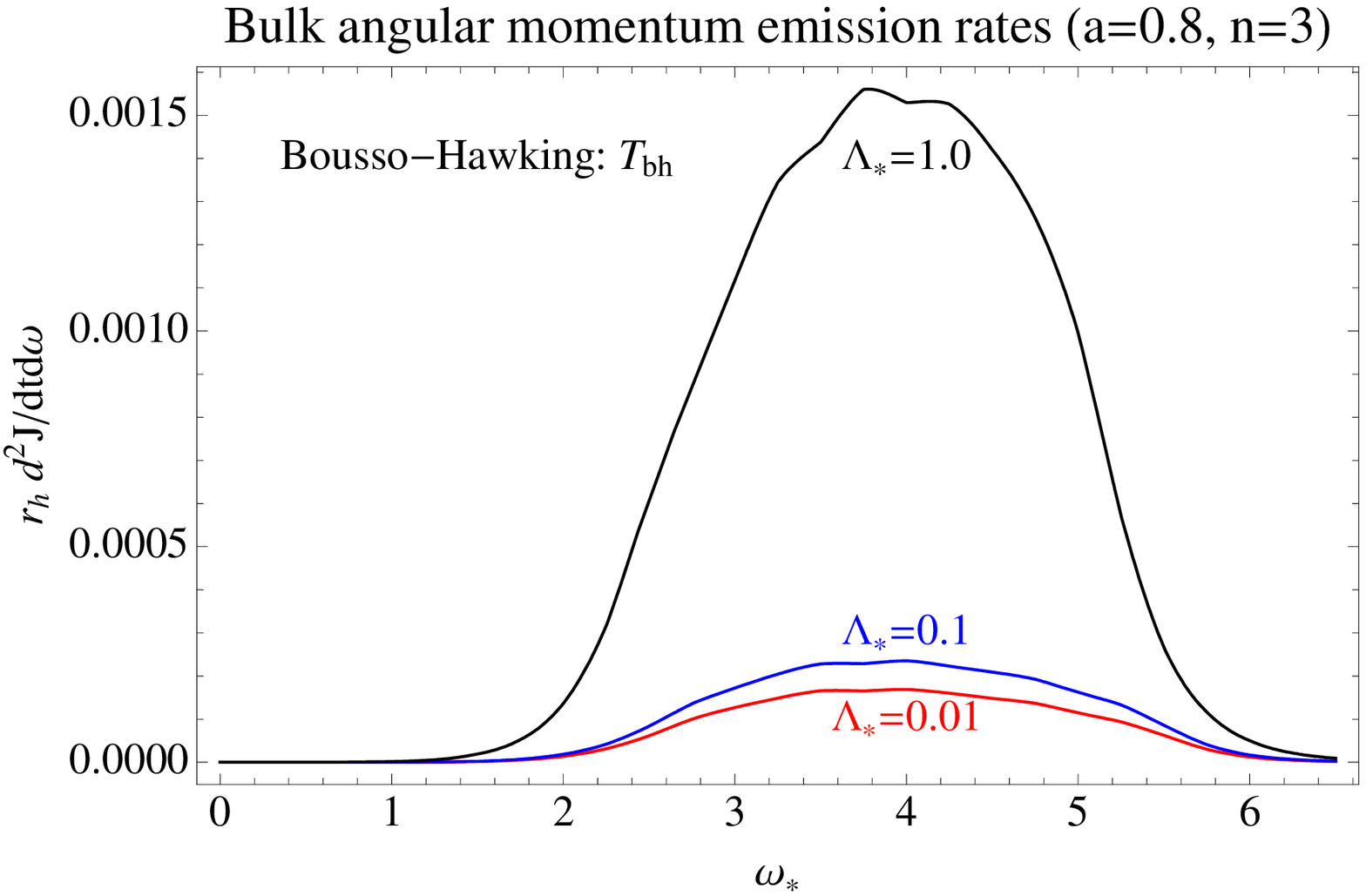}}
\vspace{-0.25cm}
\caption{\it Comparison of dS emission for Hawking and Bousso-Hawking-like temperatures for different values of $\Lambda r_h^2$.}
\label{EmissDS}
\end{figure}


\section{Conclusion}  

\par In this study the Hawking emission of the transverse traceless tensor graviton modes in asymptotically flat spacetime from higher dimensional Myers Perry black holes, see Figs. (\ref{EmissAF1}) and (\ref{EmissAF2}), were presented for rotations $a_{\star}=0.8, 1.2$,  and $a_{\star}=1.5$ respectively. This is the first time that the Hawking emission of these perturbations has been calculated, as far as the authors are aware. 

\par Perhaps the most interesting result from our investigation of the Kerr-dS case is the effect that the cosmological constant has on enhancing superradiance as is seen clearly in Fig. \ref{fig:SRCompare}. A larger cosmological constant leads to more superradiance and hence will cause the black hole to spin down more quickly. Actually, this effect and the general issue of Hawking radiation on Kerr-dS spacetime deserves more attention, with the motivation being that emissions from mini-TeV black holes in the early universe may have some observable effect on cosmology.

\par Thus, in Fig. (\ref{EmissDS}), we have also attempted to compare graviton emission in Kerr de Sitter space (for $a_\star=0.8$ and $n=3$) by working with two normalizations for the surface gravity: the normal Hawking temperature (valid for Kerr-AF) and a Bousso-Hawking-like temperature (valid for S-dS). As can been seen from these plots the larger the cosmological constant the larger the amount of radiation for the Bousso-Hawking case (in line with results found in references \cite{Kanti:2005ja,Wu:2008rb}), while the usual Hawking normalization leads to small decreases. However, as discussed in references \cite{Myung:2007my,Wu:2008rb} the use of the Bousso-Hawking temperature might lead to ill-defined black hole entropies. It would be interesting to investigate in more detail the issue of the correct normalization of the surface gravity for Kerr-dS black holes.

\par We cannot truly comment on the issue of whether or not black holes radiate mainly on the brane for simply rotating Kerr-dS black holes until the full graviton emissions are known (there are also the contributions from the scalar and vector modes). However, as pointed out in reference \cite{Cardoso:2005mh}, of the degeneracy factors for each respective perturbation: scalar ($D_j^S$), vector ($D_j^V$) and tensor ($D_j^T$), the tensor modes start to dominate as the number of dimensions increases. Thus for $n$ large we would expect the tensor mode emissions to dominate. 

\par Hopefully, within this decade, a separable set of Master equations for all the graviton perturbations will be obtained. Despite not having this set of equations there is still much work that remains to be done on the Kerr-(A)dS backgrounds. Although brane dominated emission is the standard consensus, if more realistic models of braneworld black holes are considered, such as split fermion models, which have other fields besides the graviton in the bulk, see reference \cite{Cho:2007de}, then bulk emissions may dominate. Thus, it would be interesting to examine the Hawking emission of fermions on Kerr-dS backgrounds, along the lines presented in this paper, as a generalisation of previous works for bulk fermion perturbations on Schwarzchild \cite{Cho:2007de} and tense brane \cite{Cho:2008xq} metrics.


\acknowledgments

\par HTC was supported in part by the National Science Council of the Republic of China under the Grant NSC 96-2112-M-032-006-MY3. Financial support from the Australian Research Council via its support for the Centre of Excellence for Mathematics and Statistics of complex systems is gratefully acknowledged by J.~Doukas.


\end{document}